\documentclass{aa}
\usepackage{CJK}
\usepackage[colorlinks=true, linkcolor=blue, citecolor=blue, filecolor=blue,urlcolor=blue]{hyperref}
\usepackage{graphicx,natbib,txfonts,color,xspace,amsmath,mathtools,afterpage}

\bibpunct{(}{)}{;}{a}{}{,}

\begin{document}
\begin{CJK}{UTF8}{mj}
        \title{Power density spectra morphologies of 
        seismically unresolved red-giant asteroseismic binaries}
        \author{Jeong Yun Choi\inst{1,2} \and Francisca Espinoza-Rojas\inst{1,2}  
                \and Quentin Copp{\'e}e\inst{1,2} \and Saskia Hekker\inst{1,2}}
        \institute{Heidelberg Institute for Theoretical Studies (HITS), Schloss-Wolfsbrunnenweg 35, D-69118 Heidelberg, Germany\\ 
        \email{jeongyun.choi@h-its.org} 
        \and 
        Heidelberg University, Centre for Astronomy, Landessternwarte, K\"onigstuhl 12, D-69117 Heidelberg, Germany} 
        \authorrunning{Jeong Yun Choi}
        \titlerunning{Power density spectra morphologies of 
        seismically unresolved red-giant asteroseismic binaries}
        
        \abstract{Asteroseismic binaries are two oscillating stars detected in a single light curve. These systems provide robust constraints on stellar models from the combination of dynamical and asteroseismical stellar parameters. Predictions suggested that approximately 200 asteroseismic binaries may exist among the \textit{Kepler} long-cadence data, and the majority of them consist of two red-clump (core helium burning) stars. However, detecting these systems is challenging when the binary components exhibit oscillations at similar frequencies that are indistinguishable (i.e., unresolved asteroseismic binaries).}
        {In this study, we predict the morphologies of power density spectra (PDS) of seismically unresolved red-giant asteroseismic binaries to provide examples which can be used to identify the systems among observed stars.}
        {We created 5,000 artificial asteroseismic binary (AAB) systems by combining the KASOC light curves of red giants with oscillations at similar frequency ranges. To quantify the complexity of the oscillation patterns, we used the maximum signal-to-noise ratio of the background-normalized PDS and Shannon entropy. Additionally, we identified the radial and quadrupole mode pairs for the individual binary components and determined their impact on the PDS morphologies of AABs.}
        {Our results reveal that the majority of AABs ($\sim$47\%) consist of the two red-clump stars. The PDS of AABs generally exhibits increased Shannon entropy and decreased oscillation power compared to individual components. We focused on the $\sim$8\% of AABs whose stellar components have similar brightness and classified them into four distinct morphologies: (i) single star-like PDS, where oscillations from one component dominate, (ii) aligned, where the dominant oscillations in the stars that form the AAB appear at similar frequencies, (iii) partially aligned, where some oscillation modes of component stars are aligned while others are not, and (iv) PDS containing complex structures with unclear mode patterns caused by the misalignment of the mode frequencies of both components.}
        {We found that most AABs with detectable oscillations from both components show complex oscillation patterns. Therefore, unresolved asteroseismic binaries with low oscillation power and complex oscillation patterns as characterized by high Shannon entropy offer a potential explanation to understand the observed stars with complex PDS.}
        \maketitle

\section{Introduction}  
In the universe, binary stars are as abundant as single stars \citep{1983_Abt, 2010_Raghavan, 2013_Duchene_Kraus, 2017_Moe_DiStefano}. Both binary system components share a common formation history, resulting in the same initial chemical composition and age \citep{2018_Prsa}. This allows us to constrain input parameters for stellar models and study diverse evolutionary paths depending on mass, luminosity, and their interaction \citep{2012_Sana, 2023_Tauris, 2024_Marchant}. Additionally, radial velocities from spectroscopic observations and light curve analysis of eclipsing system enabled us to directly measure the masses and radii of the binary components through dynamical modelling \citep{1991_Andersen,2010_Torres}. 

Asteroseismology has emerged as a potent method for inferring stellar parameters through scaling relations \citep{1991_Brown, 1995_Kjeldsen, 2010_Huber, 2013_Mosser, 2020_Hekker}, especially for stars with a convective surface layer. These stars show solar-like oscillations, which are intrinsically damped and stochastically excited by the turbulence in the outer layers of the stars. Accordingly, observing these oscillations allowed us to explore the internal structure of the stars. Using the high-precision long-term time series data from the \textit{Kepler} space mission \citep{2010_Borucki}, solar-like oscillations have been detected in about 20,000 red giants \citep{2011_Bedding, 2011_Hekker, 2011_Huber, 2014_Pinsonneault, 2018_Pinsonneault, 2024_Pinsonneault, 2018_Yu}.
 
Among many identified binary systems with oscillating components (see reviews by \citealt{2018_Murphy_review, 2024_Murphy}), solar-like oscillators in these binary systems are particularly valuable. This is because the dynamical masses derived from orbital analysis can provide constraints on the asteroseismic scaling relation and further improve stellar modeling (e.g., \citealt{2010_Hekker, 2013_Gaulme, 2014_Beck, 2018_Beck, 2022_Beck, 2024_Beck, 2018_Li_Tanda, 2018_Themessl}, and references therein). If both components exhibit oscillations, they provide constraints for both stars and the full potential of asteroseismology. Especially, when oscillations from two stars are detected in a single light curve, we refer to them as asteroseismic binaries, regardless of whether the two components are gravitationally bounded or not. However, only a few asteroseismic binaries are discovered in which both stars exhibit solar-like oscillations. \citet{2015_Appourchaux} reported HD 177412 (KIC 7510397) with solar-like oscillations in two main-sequence binary stars and compared the astrometrical measurement of stellar parameters to the seismic analysis. Moreover, \citet{2018_Marcadon} studied triple star system HD 188753 (KIC 6469154) where they detected solar-like oscillations in the two brightest components which have oscillation modes around 2,200 and 3,300 $\rm{\mu}$Hz in the power density spectra (PDS), respectively. KIC 9163796 has been studied by \citet{2018_Beck}, and recently by \citet{2025_Grossmann}, where the power excess of the secondary component is above the Nyquist frequency. For red giants, \citet{2018_Themessl} investigated KIC 2568888 which shows the asteroseismic signals of two red-giant components. Further investigations of red-giant asteroseismic binaries in wide binary systems will be presented by Espinoza-Rojas et al. (in prep.).

While asteroseismic binaries with stars having oscillations in similar frequency ranges are challenging to discover, a handful of systems have been reported. \citet{2017_White} presented the main sequence binary system HD 176465 (KIC 10124866) and derived stellar parameters by analyzing seismic signals of both components, together with an orbital analysis. \citet{2018_Li} analyzed KIC 7107778 which is a non-eclipsing and unresolved asteroseismic sub-giant binary system with oscillations of both components completely overlap. Even though one of two solar-like oscillations is marginally detectable, KIC 9246715 was identified as an eclipsing red-giant binary containing two nearly identical red giants where both components exhibit oscillations at similar frequency ranges (see \citealt{2016_Rawls}). Only with these few examples, the characteristics of red-giant asteroseismic binaries with overlapping oscillation patterns remain unexplored.
 
\cite{2014_Miglio} conducted population studies on non-eclipsing asteroseismic binaries, predicting that 200 or more such systems should be detectable in \textit{Kepler} long-cadence data. Their study concluded that combinations of red giants are most prevalent among asteroseismic binaries which have solar-like oscillations of two stars in the frequency spectrum of a single lightcurve. Most of these binaries are expected to consist of two core-He-burning (CHeB) red giants. Moreover, a recent study by \cite{2025_Mazzi} confirmed that binaries consisting of two CHeB show the largest fraction of systems detected as asteroseismic binaries in their simulation. Especially, red clump (RC) stars that have gone through a helium flash typically have similar core masses and luminosities. Thus, all RC stars exhibit oscillations in a similar frequency range with their frequency of maximum oscillation power ($\rm{\nu_{max}}$) typically around ${\rm{20-50}}$ $\rm{\mu}$Hz. If two stars with similar $\rm{\nu_{max}}$ are closer to each other in the plane of the sky than the spatial resolution of the telescope, their oscillation signals overlap in the PDS. In this case, their oscillation patterns may look different from single stars. Furthermore, the additional light from the presence of the other star can cause photometric dilution which reduces the relative flux variations, making the analysis of these systems more difficult. 

In this work, we created 5,000 seismically unresolved artificial asteroseismic binary (AAB) systems to investigate how they appear in the PDS. Each AAB is composed of two red giants which exhibit oscillations at similar frequency ranges. The resulting variety of PDS morphologies can be considered as a template for identifying potential asteroseismic binary candidates in the observed data. Moreover, we used the Shannon entropy as an estimator to quantify the complexity in the shape of PDS. 

Claude Shannon first introduced entropy in communication theory (\citealt{1948_Shannon}). Over the recent decades, Shannon entropy has been widely applied in the fields of information transmission, computer sciences, astronomy, etc. In astronomy, \citet{1995_Cincotta, 1999_Cincotta} used Shannon entropy to search for periodicity in astronomical time series. Moreover in asteroseismology, the studies by \citet{2021_Audenaert} and \citet{ 2022_Audenaert} used multiscale entropy to capture the complexity of time series at different timescales and classify different types of stellar variability. Furthermore, \citet{2022_Suarez} applied Shannon entropy to detect frequency patterns in the PDS of $\delta$ Scuti stars, determining the large frequency separation when the entropy in the {\'e}chelle diagram reaches its minimum. Building on these insights, we use the Shannon entropy to provide a measure of the complexity of the PDS with signal-to-noise ratio (SNR) to investigate different PDS morphologies. Additionally, we focused on the radial ($l = 0$) and quadrupole ($l = 2$) oscillation modes of binary components, as they offer the insight into the structure of the PDS.

The paper is organized as follows. In Sect. \ref{sec:Data} we discuss the timeseries data of single stars that we used to form AABs. In Sect. \ref{sec:Method} we describe the analysis of timeseries data and power density spectra. We also introduce the application of Shannon entropy and how we create AABs. In Sect. \ref{sec:Results} we present the analysis of 5,000 AABs and examine how the oscillation patterns and power of AABs differ from those of the single stars. Then, we classify the morphologies of AABs. Finally, we discuss the results and conclude in Sect. \ref{sec:Conclusion}.

\section{Timeseries data} \label{sec:Data} 
We used red giant stars from the second APOKASC catalog (APOKASC2; \citealt{2018_Pinsonneault}) which contains 6676 evolved stars. The APOKASC project combined \textit{Kepler} asteroseismic data with APOGEE spectroscopic data to obtain precise determination of stellar properties for a large sample of red giants. We used \textit{Kepler} long-cadence (LC; 29.45 min sampling) pre-processed light curves available from the \textit{Kepler} Asteroseismic Science Operations Center (KASOC) database\footnote{\href{http://www.kasoc.phys.au.dk/}{www.kasoc.phys.au.dk}}. All light curves used in this work were corrected using the KASOC pipeline (see \citealt{2014_Handberg_Lund} for further details). 

We selected stars which are observed for more than 1000 days and have a filling factor of 0.8 or higher, indicating that each star was monitored for more than 80\% of the total observation period. Additionally, we used stars with $\nu_{\text{max}}$ $\geq$15 $\mu$Hz, corresponding to stars with surface gravity (log g) larger than approximately 2.0 dex. We set the upper limit of $\nu_{\text{max}}$ to 200 $\mu$Hz, so that power excesses are well below the Nyquist frequency ($\rm{\approx}$ 283 $\mu$Hz for \textit{Kepler} long-cadence data).

Finally, we assume that the time series contain oscillations from only a single star. Therefore, we selected stars which are not included in the Kepler Eclipsing Binary Catalog KEBC\footnote{\href{http://keplerebs.villanova.edu/}{http://keplerebs.villanova.edu/}} (\citealt{2011_Prsa}; \citealt{2011_Slawson}; \citealt{2016_Kirk}), not among the oscillating red giant stars in binary systems analyzed by \citet{2022_Beck, 2024_Beck}, and not on the list of confirmed chance alignment and contact binary stars from \citet{2017_Colman}. We also checked non-single stars from \textit{Gaia} Data Release 3 (DR3) (\citealt{2023_Gaia}). Then we focused on red giant stars whose evolutionary stages were determined by \citet{2019_Elsworth} based on four different studies. The final dataset consists of 3,063 stars which contains 1,713 red-giant branch (RGB) stars, 1,180 RC stars, and 170 secondary clump (2CL) stars. 

\section{Methods} \label{sec:Method}
In this section, we describe how we fit the background model to normalize power density spectra (PDS) in Sect. \ref{subsubsec:Background} and estimate $\rm{\nu_{max}}$ in Sect. \ref{subsubsec:Methods_numax}. Since we focus on seismically unresolved artificial asteroseismic binary (AAB) systems, we applied the same background modeling procedure to both single stars and AABs. To compare the oscillation patterns between individual components which affect the PDS morphologies of AABs, we identified oscillation modes and estimated $\rm{\Delta\nu}$ in the PDS of the individual stars, as described in Sect. \ref{modeID}. We introduce the application of Shannon entropy to investigate the complexity of PDS in Sect. \ref{subsec:entropy} and the procedure of creating AABs in Sect. \ref{subsec:create_PDS}.

\subsection{Asteroseismic analysis} \label{subsec:Method_Analysis}

\subsubsection{The background model} \label{subsubsec:Background}
In the Fourier spectrum, power from stellar activity, granulation, and noise (i.e. photon noise and instrumental noise) creates a background on which oscillations are superimposed. \citet{2014_Kallinger} found that the instrumental noise is negligible, and thus the noise can be assumed to be white noise. Normalizing these background signals is necessary to study oscillations. We used a prototype of the TACO (Tools for the Automated Characterisation of Oscillations, Hekker et al., in prep.) code to fit a background and obtain $\rm{\nu_{max}}$. Following the description by \citet{2014_Kallinger}, the global background fit of the PDS ($\rm{P_{bg}(\nu)}$) is modeled as 
\begin{equation}
\rm{P}_{bg}(\nu) = \it{w}_{{\textrm{noise}}} + \rm{\eta}^2(\nu)\left(\sum_{i=1}^{3}\frac{A_i}{1+(\nu/b_i)^4}\right),   \label{eq:background}
\end{equation}
with $w_{\textrm{noise}}$ as the white noise, and the sum of three super-Lorentzian profiles represent granulation components occurring at different length- and time-scales with their characteristic amplitude $\rm{A_i}$ and characteristic (turnover) frequency $\rm{b_i}$ (blue dashed lines in Fig.\ref{fig:Figure-synthetic}). The granulation signals are affected by \textit{apodization} $\rm{\eta(\nu)}$ due to the discrete sampling of the flux measurements (see \citealt{2014_Kallinger} for details). 

\subsubsection{$\rm{\nu_{max}}$ estimation} \label{subsubsec:Methods_numax}
Oscillations are visible as a power excess in the PDS, and we can fit a gaussian functions to estimate $\rm{\nu_{max}}$, the frequency of maximum oscillation power. The power excess is modeled as:
\begin{equation}
\rm{P}_{{osc}}(\nu) = P_{\rm{g}}\exp \left(- \frac{(\nu-\nu_{\rm max})^2}{2\sigma_\mathrm{env}^2}\right)
\label{eq:oscillation}
\end{equation}
where $\rm{P}_{\rm g}$ represents the height of the Gaussian envelope and $\rm{\sigma_{env}}$ is the width of the Gaussian envelope. Therefore, the full PDS is modeled as 
\begin{equation}
\rm{P(\nu) = P_{bg}(\nu) + \eta^2(\nu)P_{osc}(\nu)}.
\label{eq:global-fit}
\end{equation}
To estimate parameters in equation (\ref{eq:global-fit}), we applied \textit{Bayesian Markov Chain Monte Carlo} (MCMC) framework with affine-invariant ensemble sampling (emcee; \citet{2013_Foreman_emcee}). Then we obtained the posterior probability distributions for each parameter and we adopted the medians of these distributions as an estimate of the expectation values for the parameters and their 16th and 84th percentiles as standard uncertainties. With the global background model, we can create background-corrected PDS (i.e. the observed PDS $\rm{P_{obs}(\nu)}$ divided by $\rm{P_{bg}(\nu)}$), see Fig. \ref{fig:Figure-synthetic}. 

\subsubsection{Identification of the \textit{l} = 0, 2 modes, and $\Delta\nu$ estimation}  \label{modeID}
We applied the peak-finding method described by \citet{2018_Garcia} to identify statistically-significant Lorentzian-like peaks in the background-normalized PDS. Each detected peak was fitted using a Lorentzian function given by:
\[\rm{
P_{\rm peak}(\nu) = \frac{H_{\rm peak}}{1 + \left(\frac{\nu - \nu_{\rm peak}}{\gamma_{\rm peak}}\right)^2},
}\]
where \(\nu_{\rm peak}\), \(H_{\rm peak}\), and \(\gamma_{\rm peak}\) represent the central frequency, height, and half-width at half maximum (HWHM), respectively. We optimized the parameters of each detected peak using Maximum Likelihood Estimation (MLE), with a lower limit of the uncertainties calculated by using the Hessian matrix.

To identify the radial ($l=0$) and quadrupole ($l=2$) modes, we used the “universal pattern” \citep{2011_Mosser} for solar-like oscillations, which predicts the expected mode frequencies. We cross-correlated the observed PDS with the universal pattern within the frequency range $\nu_{\mathrm{max}}\pm\Delta\nu$ to locate the central $l=0$ mode closest to $\nu_{\mathrm{max}}$. By using the identified central $l=0$ frequencies, we calculated the corresponding $l=2$ modes for the given radial order. Subsequently, we identified $l=0,2$ modes further away from $\nu_{\mathrm{max}}$. Finally, we calculated the large frequency spacing ($\Delta\nu$, the frequency difference between modes of the same spherical degree and consecutive radial orders) from a weighted linear fit to the identified $l=0$ modes, considering their frequency uncertainties. 

\subsection{Shannon Entropy} \label{subsec:entropy}
\begin{figure*}[!htbp]
    \centering
    \includegraphics[width=15.5cm, height=22cm]{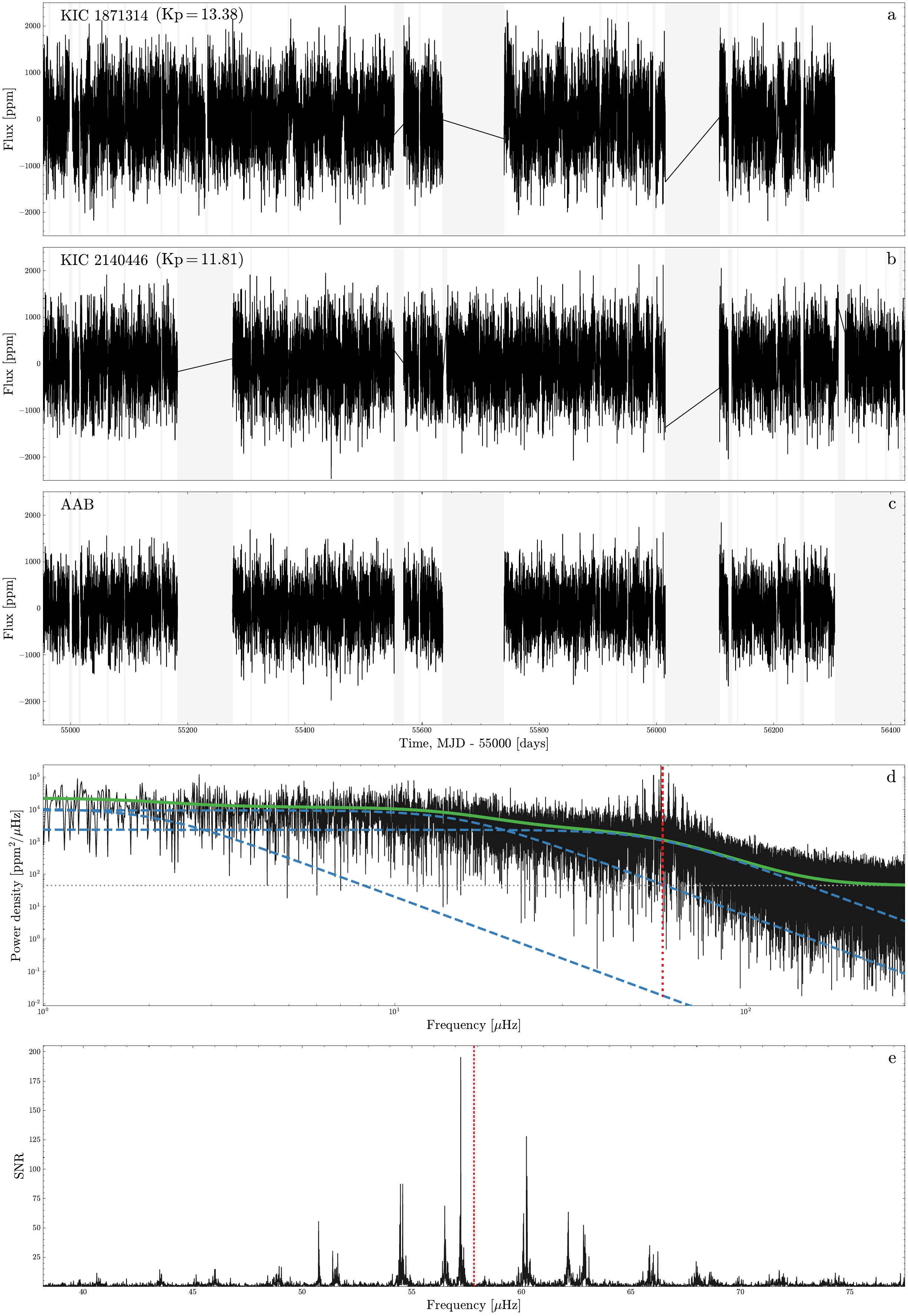}
    \caption{Process of creating the light curve of an artificial asteroseismic binary star (AAB) and its power density spectra. Panels a and b: Individual light curves of two red giant stars (KIC 1871314 and KIC 2140446). Panel c: The combined artificial light curve. The shaded light-gray regions indicate areas where data points are not available. Panel d: Power density spectrum of the light curve from panel c. In this panel, the blue dashed lines represent granulation background components, the gray dotted line indicates the white noise component, and the green line shows the total background fit. Panel e: Background-corrected power density spectrum in the frequency range of $\rm{\nu_{max} \pm 3 \sigma_{env}}$. The vertical red dashed line in panels d and e indicates the location of $\rm{\nu_{max}}$ estimated as explained in Sect. \ref{subsubsec:Methods_numax}}.
    \label{fig:Figure-synthetic}
\end{figure*}

\begin{figure*}[!htbp]
    \centering
    \includegraphics[width=1\linewidth]{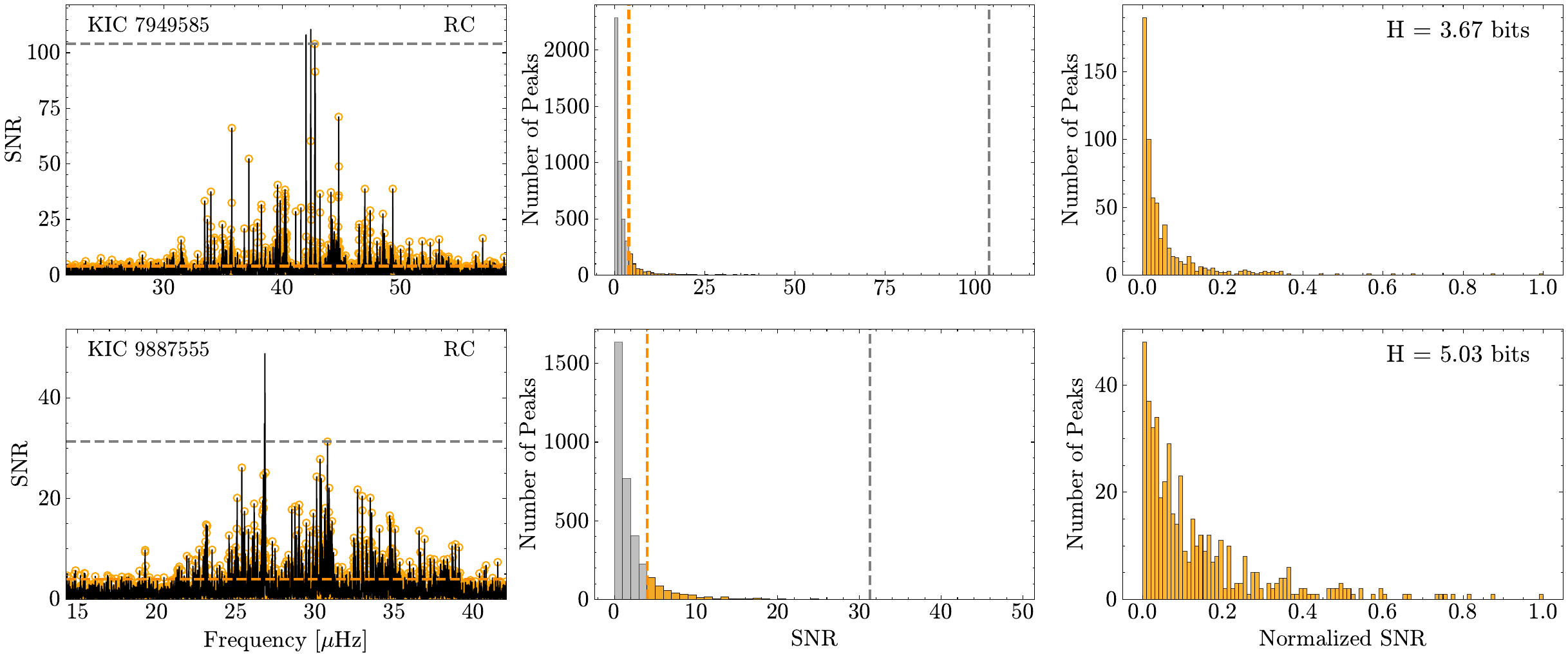}
    \vspace{0.1cm}
    \caption{Power density spectra (PDS) and corresponding histograms of power for two stars with different levels of complexity. The upper panel shows KIC 7949585, which exhibits lower entropy (low complexity), while the lower panel shows KIC 9887555, which has higher entropy (high complexity). The yellow dashed line is the SNR threshold of 4, and the gray dashed line shows maximum SNR of the selected peaks (see detail about how we select peaks in Sect. \ref{subsec:entropy}). Peaks that are used in the entropy calculation are marked as yellow circles at the peak maxima position in the left panels which fall into the yellow bins in the middle panels. In the middle panels, the histograms have a bin width of 1. We normalized their SNR to scale the values to the range from 0 to 1 (right panels). The right panels show histograms created by binning the normalized power values into 100 equally spaced bins. These are the probability distribution used in the entropy calculation Eq.(\ref{eq:shannon_entropy}).}
    \label{fig:Figure-entropy}
\end{figure*}

\begin{figure*}[!htbp]
    \centering
    \includegraphics[width=\linewidth]{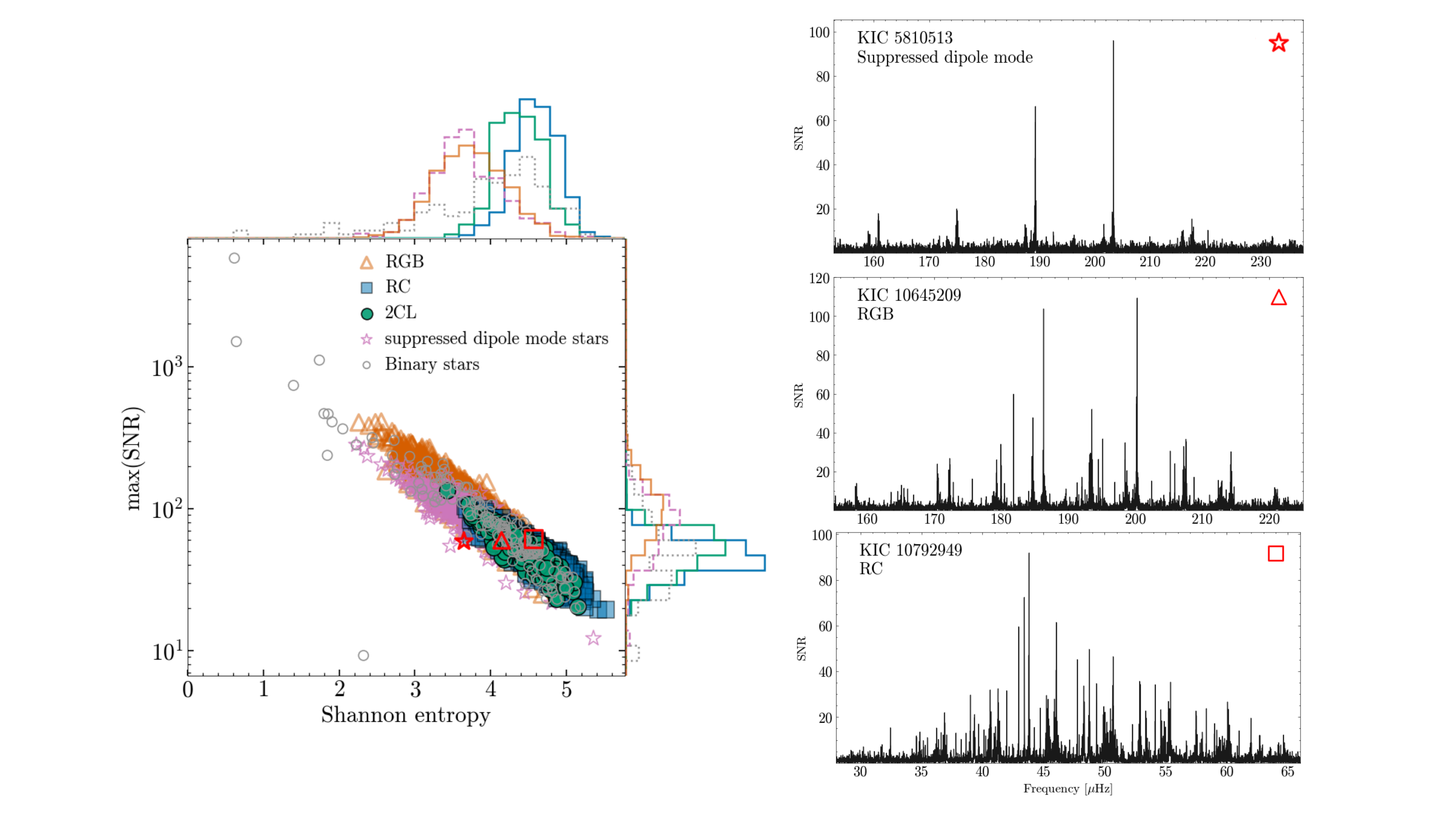}
    \vspace{0.1cm}
    \caption{Maximum SNR versus entropy of the background-normalized PDS for 3,212 observed red giant stars, with the y-axis plotted on a logarithmic scale. Known binary stars are shown as gray circle (see Sect. \ref{sec:Data} for more details of this set of stars). The panels on the right show the background-normalized PDS corresponding to specific red symbols marked on the left panel. These PDS figures highlight the characteristics associated with their different entropy levels.}
     \label{fig:Figure-Entropy_method}
\end{figure*}

We used the concept of Shannon entropy\textcolor{black}{\footnote{\textcolor{black}{Throughout this paper, we used "entropy" as shorthand for "Shannon entropy".}}} to analyze the PDS of both individual stars and artificial binaries. The entropy measures both the unpredictability in a probabilistic distribution and the amount of information it provides. \cite{1948_Shannon} defined the entropy H of a discrete random variable X as 
\begin{equation}
    {\rm{H(X) = -\sum_{\it{i}\rm{=1}}^N p(\it{x_i}\rm{)}\log p(\it{x_i}\rm{)}}},  \label{eq:shannon_entropy}
\end{equation}
where X is a discrete random variable that takes values from the finite set \( \{x_1, x_2, \dots, x_N\} \) with corresponding probabilities given by the probability distribution function $\rm{p(X = \it{x_i}\rm{)}}$. The choice of base for logarithm is typically taken to base 2, which gives the unit of bits. 

We estimated $\rm{p(\it{x_i}})$ using a histogram derived from the observed power values in the PDS. For modes with large linewidth, multiple adjacent peaks are selected as part of a single mode, which could slightly increase the estimated entropy. However, this effect is not significant as the spread in the mode linewidths is not large. Then, we normalized the PDS by the global background fit (Eq. \ref{eq:background}), which provides the power density values equivalent to the signal-to-noise ratio (SNR). We focused on the frequency range of $\rm{\nu_{max} \pm 3 \sigma_{env}}$ (see the bottom panel of Fig. \ref{fig:Figure-synthetic} and the left panels of Fig. \ref{fig:Figure-entropy}). 

A threshold value of SNR is set to reduce the influence of noise peaks in the background-normalized PDS. We empirically chose an SNR threshold of 4 and consider only those SNR values that are 4 or above. Additionally, to mitigate the effect of occasional spikes with unusually high SNR, we removed one or two highest peaks if the SNR is above 99.95th percentile of the SNR distribution.  

Then we normalized the selected power values (i.e., only those with SNR $\geq$ 4) so that they range from 0 to 1, by the following transformation: 
\begin{equation}
\rm{SNR_{normalized}=\frac{SNR-4}{max(SNR)-4}}
\label{eq:shannon_entropy_norm}
\end{equation}
where SNR indicates the background normalized power density values in the selected frequency range and max(SNR) represents the highest value among these SNRs. We created a histogram of the normalized SNR values using 100 bins uniformly distributed over the range [0,1] (see the right panels of Fig. \ref{fig:Figure-entropy}), each bin is labeled as $\it{x_i}$ (where \textit{i }in \{1, 2, \dots, 100\}). We used this histogram to estimate the probability distribution $\rm{p}({\it{x_i}}\rm{)}$. For each bin labeled as $\it{x_i}$, we count the number of normalized SNR values that fall within that bin and denote this count as $\it{n_i}$. The total number of data points across all bins is
\begin{equation}
\rm{N}=\sum_{\it{i}\rm{=1}}^{100}n_{\it{i}}. 
\end{equation}
We estimated the probability of observing a normalized SNR around $\it{x_i}$ as  
\begin{equation}
    \rm{p}({\it{x_i}}\rm{)}=\frac{n_{\it{i}}}{\rm{N}}, 
\end{equation}
which represents the likelihood of observing a peak at a given normalized power level in the background-normalized PDS. Because the probabilities across all bins must sum to 1, we have: 
\begin{equation}
\rm{\sum_{\it{i}\rm{=1}}^{100}\rm{p}({\it{x_i}}\rm{)=1}}.
\end{equation}

In cases where max(SNR) values differ significantly between stars, differences in entropy could partly result from the normalization effect (Eq. \ref{eq:shannon_entropy_norm}) rather than intrinsic differences in oscillation complexity. To mitigate this, we filtered out peaks with extreme SNR values before normalization and used max(SNR) as a complementary parameter. Consequently, entropy robustly highlights exceptionally complex oscillation patterns that would otherwise be difficult to identify (see Fig. \ref{fig:Figure-entropy}). 

When comparing stars with similar max(SNR) values, entropy provides an even more powerful metric for distinguishing PDS with different levels of complexity. Specifically, higher entropy reflects a more uniform, less predictable distribution of oscillation powers that conveys a larger amount of information, whereas lower entropy indicates a more predictable distribution with less information content. We calculated the entropy for 3,212 observed red giant stars (3,063 from main dataset and 149 known binary stars, see Fig. \ref{fig:Figure-Entropy_method}). Notably, for stars with similar maximum SNR, the RC star with more prominent peaks exhibit higher entropy values compared to the RGB star and those with lower entropy values show a higher likelihood of being suppressed dipole mode stars (see the right panels in Fig. \ref{fig:Figure-Entropy_method}). Suppressed dipole mode stars exhibit dipole mode visibility well below expected levels (for more details, see \citealt{2024_Coppee} and earlier works from e.g. \citealt{2012_Mosser}, \citealt{2016_Stello}, and \citealt{2017_Mosser}). Additionally, binary stars exhibit a broad range of entropy values, potentially due to the peaks originating from the orbital modulation in the binary system. 

\subsection{Construction of AABs} \label{subsec:create_PDS}
The observed stellar oscillation amplitude is reduced if there is additional light coming from nearby sources within the photometric aperture. As a result, the relative amplitude of the oscillation signal will be smaller compared to the single isolated star, since the detected flux variations are diluted by the total combined light. For the case of a seismically unresolved binary system, the observed light curve reflects the combined photometric fluctuations from both stars within the total flux of the system. Therefore, we incorporated this dilution effect inherently by considering their relative brightness. To calculate the flux contribution of each star, we divided the flux of each star by the combined flux of the two stars. For instance, KIC 1871314 ($\mathrm{Kp =13.38}$ mag, where $\mathrm{Kp}$ is the magnitude in the \textit{Kepler} band) contributes approximately 19\%, while KIC 2140446 ($\mathrm{Kp =11.81}$ mag) accounts for about 81\% (see Fig.~\ref{fig:Figure-synthetic}). Then we multiplied the flux in the light curves of each star with the respective fractional contribution. 

Since stars were not observed at exactly the same epochs, we linearly interpolated the timeseries of the star with fewer observations to match the epochs of the star with more numerous observations. The interpolation is not performed across significant observational gaps. By adding the resulting interpolated flux values from both stars, we created a light curve of AAB. The entire procedure is shown in the Fig. \ref{fig:Figure-synthetic}.  

From the 3,063 stars in APOKASC2 (see Sect. \ref{sec:Data}), we can create 4,689,453 unique AABs. To focus on unresolved red-giant asteroseismic binaries that oscillate within a similar frequency range, we selected only those pairs whose difference in $\nu_{\mathrm{max}}$ is within 10\% of their average $\nu_{\mathrm{max}}$ and a filling factor of 0.8 or higher, resulting in 517,184 artificial binaries. From this subset, we randomly selected 5,000 artificial binaries.

\section{Results} \label{sec:Results}
With 5,000 AABs, we explore the various combinations of binary components in terms of their evolutionary stage, $\nu_{\mathrm{max}}$, flux ratio, entropy, and maximum SNR of background normalized PDS. We compare the background normalized PDS of AABs with those of the individual components, investigating differences in entropy and oscillation power. Finally, we classify AABs according to the morphological variations in their background normalized PDS.

\subsection{Distribution of binary components forming AABs} \label{subsec:subsec1_Distribution}

\begin{figure*}[!htbp]
    \centering
    \includegraphics[width=0.77\linewidth]{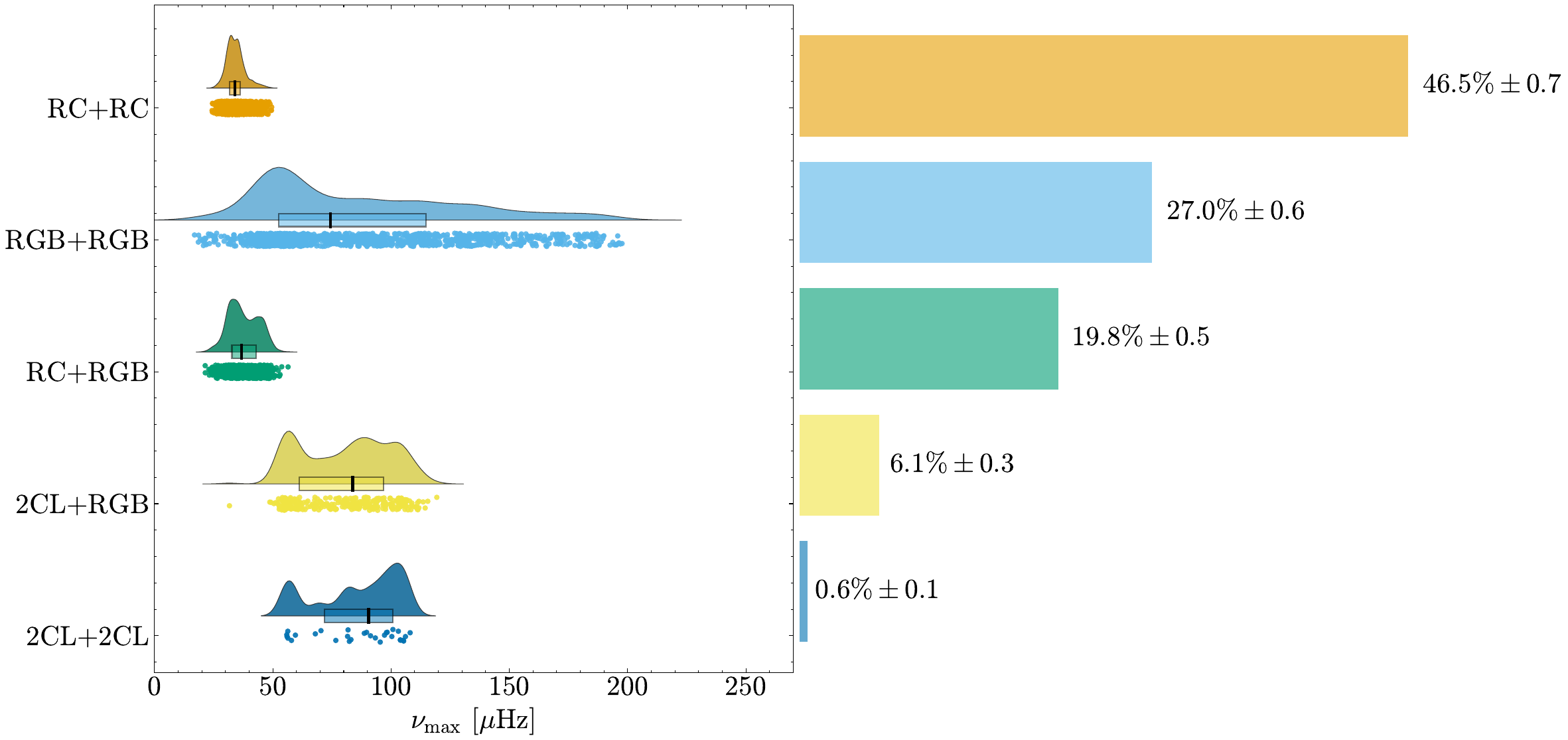}
    \caption{$\rm{\nu_{max}}$ distribution of AABs, with various evolutionary stage combinations of binary components. Below each smoothed curve (normalized to the unit area), a box shows the inter-quartile range (the 25th to 75th percentiles), with a black vertical line indicating the median. The bar chart on the right shows the percentage of different combinations with their standard deviations.}
    \label{fig:Figure-pairs}
\end{figure*}

\begin{figure}[!htbp]
    \includegraphics[width=0.95\linewidth]{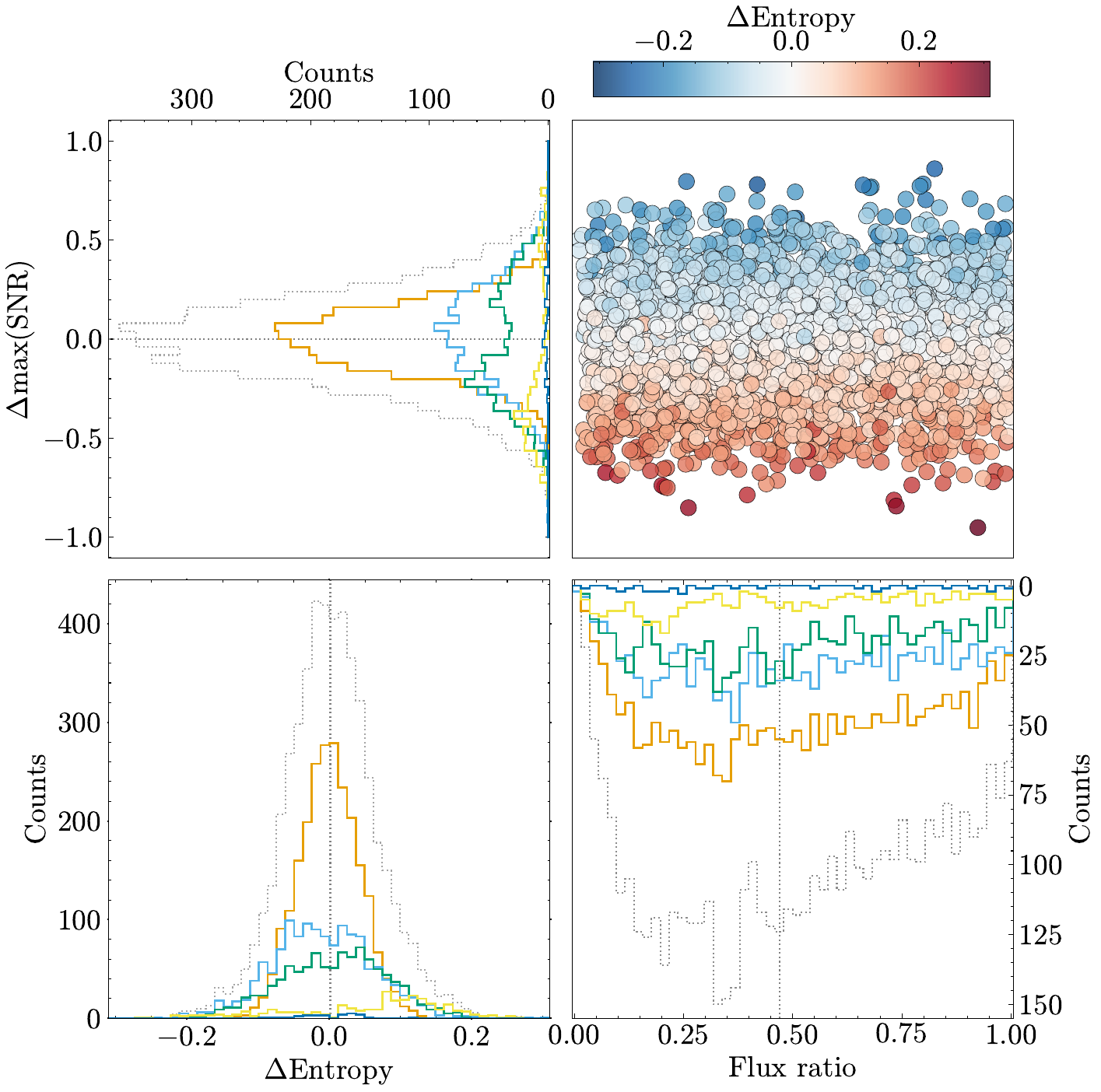}
    \caption{Binary components are characterized by flux ratio, entropy, and the maximum SNR of the background normalized PDS. The upper right panel shows the distribution of flux ratios against the maximum SNR differences, with points color-coded by entropy differences. The adjacent histograms show the respective distributions. The lower left panel shows the entropy difference distribution. The colors of the histograms have the same meaning as in Fig. \ref{fig:Figure-pairs}, with gray dotted lines representing all binary components and the vertical lines indicate the medians.}
    \label{fig:Figure-distribution}
\end{figure}

Since we created AABs by using two stars with similar $\rm{\nu_{max}}$ (see Sect. \ref{subsec:create_PDS}), combinations of RC + RC stars dominate (accounting about 47\% of the total, see Fig. \ref{fig:Figure-pairs}). The next most common is RGB + RGB ($\sim$27\%) due to the largest number of available stars selected in Sect. \ref{sec:Data}. Following this, pairs of RC + RGB stars make up about 20\%, while combinations including 2CL stars are less frequent. We calculated the standard deviation of these fractions via bootstrap resampling method, by randomly generating 1,000 samples of 5,000 AABs. We confirmed that the overall distribution remains consistent to the one provided in Fig. \ref{fig:Figure-pairs}. While our specific results are restricted to the selected stars in Sect. \ref{sec:Data}, we expect that unresolved red-giant asteroseismic binaries which are not identified yet are most likely composed of RC + RC stars, consistent with the predictions by \citet{2014_Miglio}.  

Binary components vary in brightness, maximum SNR of the background normalized PDS, and entropy. We found that the flux ratios of binary components used to create the set of 5,000 AABs are uniformly distributed, with a median around 0.5 (see Fig. \ref{fig:Figure-distribution}). We calculated the relative differences in entropy and maximum SNR ($\rm{\Delta = ({b - f})/({b + f})}$, where b and f indicate the brighter and fainter stars, respectively). The histograms for relative entropy and maximum SNR differences peak around zero, which indicates that many binary components have similar values in these parameters. This is due to the significant number of the pairs are RC + RC (see Fig. \ref{fig:Figure-pairs}), as RC stars have a relatively narrow range of max(SNR) and entropy values (see Fig. \ref{fig:Figure-Entropy_method}). In contrast, binary components involving RGB stars show a more gradual relative difference distribution both in max(SNR) and entropy. Moreover, we observed that larger differences in max(SNR) typically correspond to larger differences in entropy. Distributions of these parameters confirm that randomly generated binary components cover a comprehensive range forming a diverse set of AABs. 

\subsection{Variations in entropy and SNR between binary components and AABs} \label{subsec:subsec2_Entropy} 

\begin{figure*}[!htbp]
    \centering
    \begin{minipage}{0.32\textwidth}
        \centering
        \includegraphics[width=\linewidth]{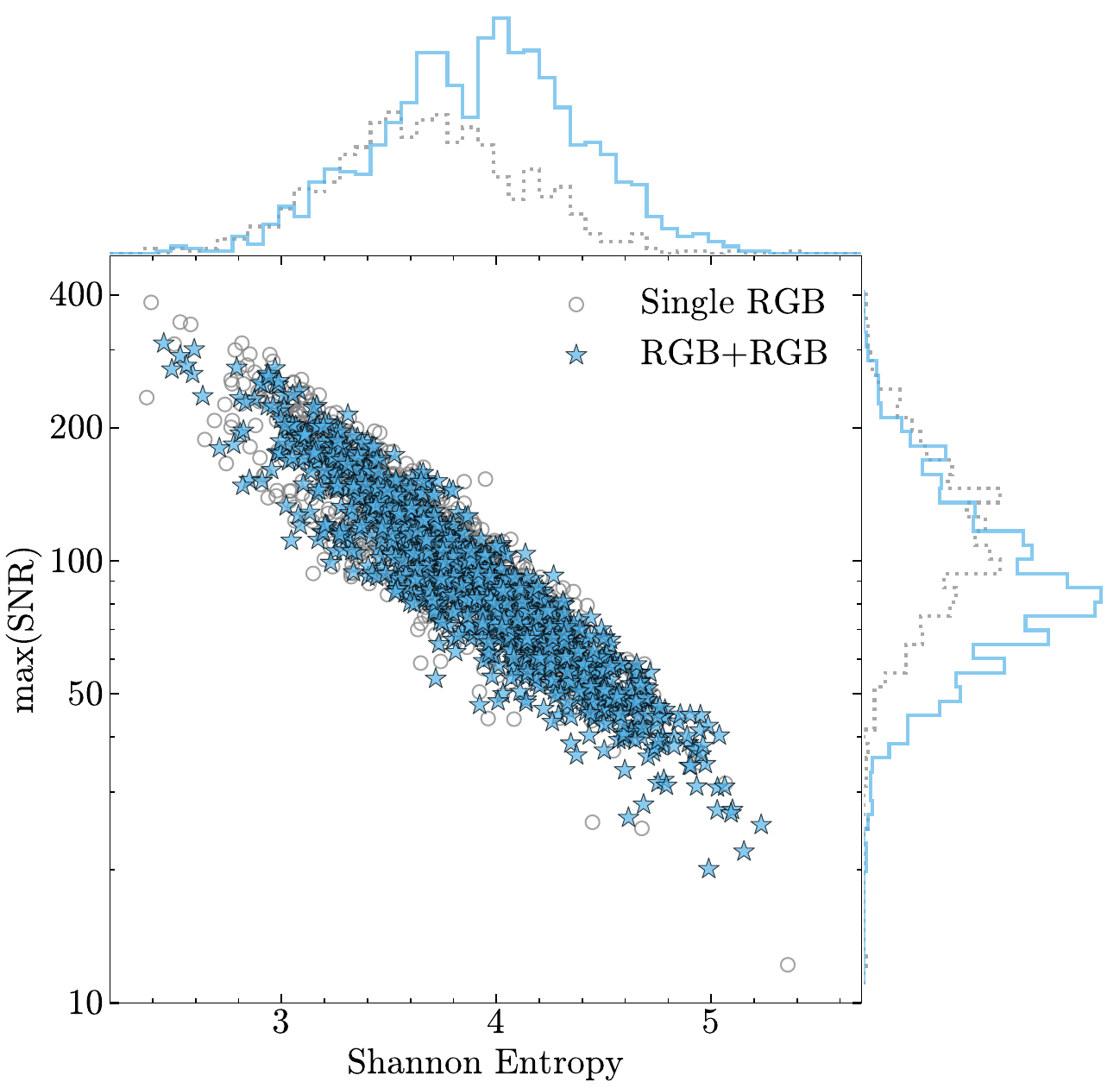}
        \label{fig:RGB+RGB}
    \end{minipage}
    \hspace{0.005\textwidth}
    \begin{minipage}{0.32\textwidth}
        \centering
        \includegraphics[width=\linewidth]{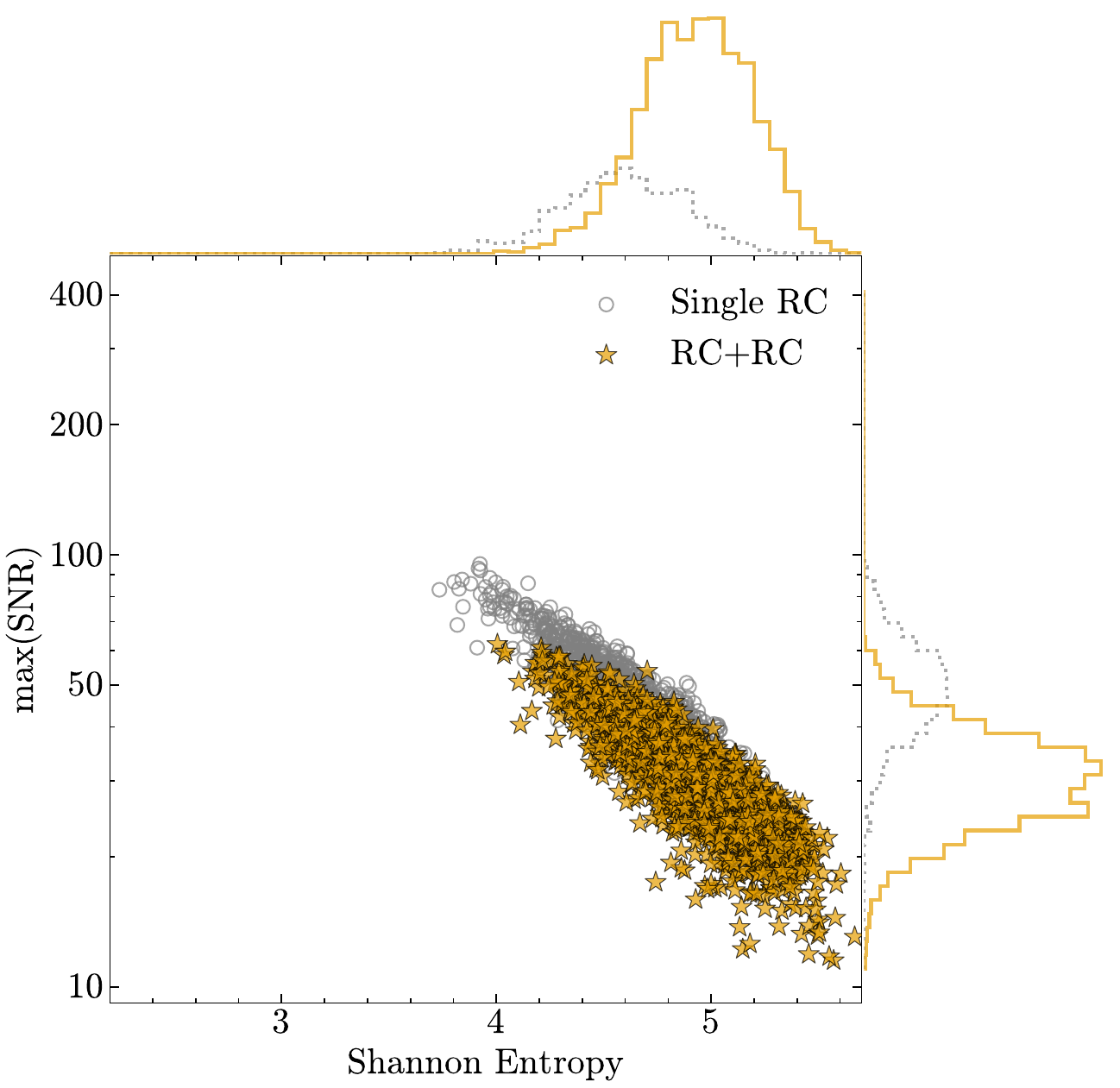}
        \label{fig:RC+RC}
    \end{minipage}
    \hspace{0.005\textwidth}
    \begin{minipage}{0.32\textwidth}
        \centering
        \includegraphics[width=\linewidth]{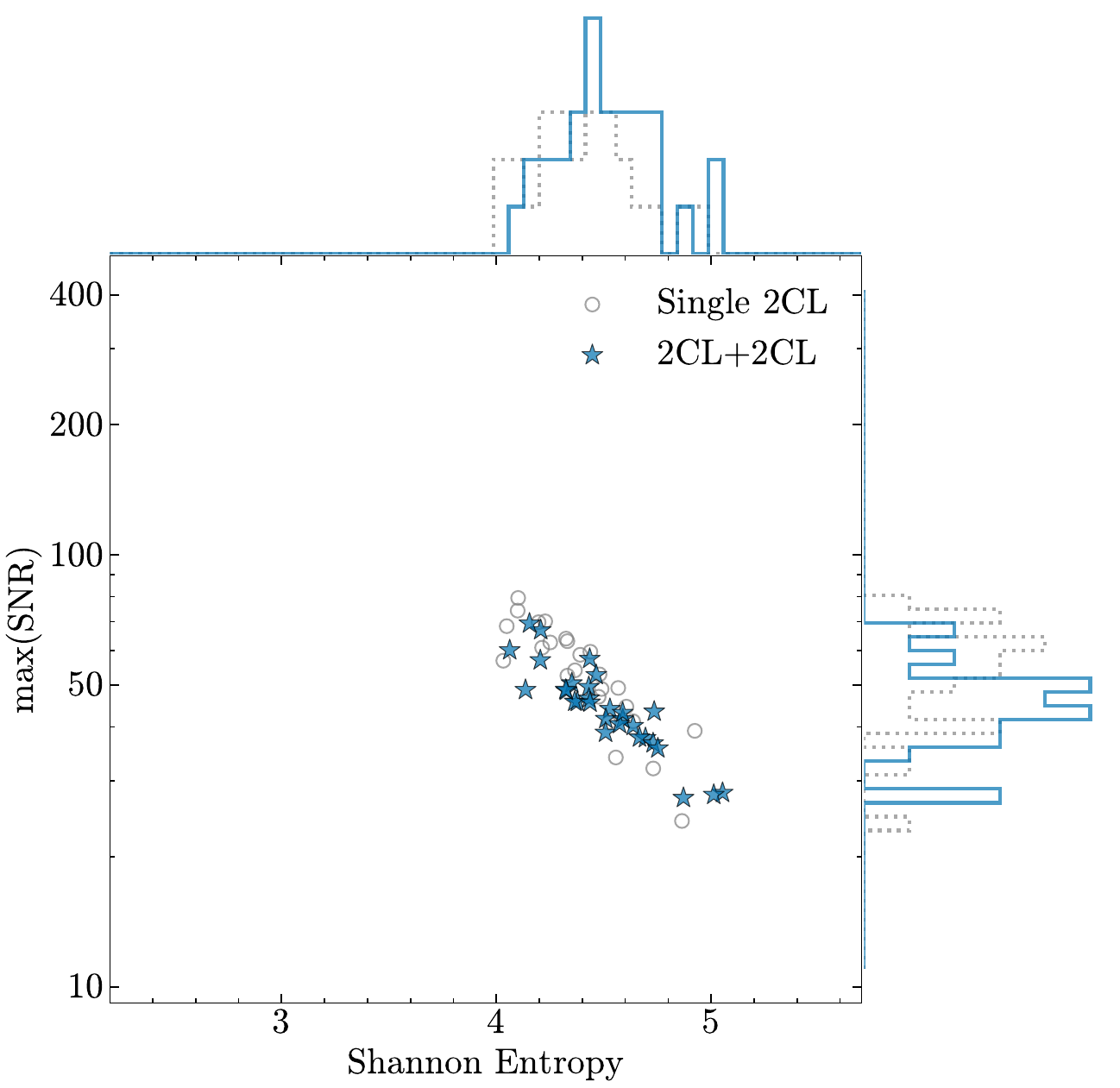}
        \label{fig:2CL+2CL}
    \end{minipage}
    \vspace{-0.5em}
    \begin{minipage}{0.32\textwidth}
        \centering
  \caption{
        Comparison of maximum SNR (logarithmic scale) and entropy between AABs (star symbols) and their component stars. Histograms along each axis represent the raw counts, with grey dashed lines representing the distributions of all single stars in each panel.}
        \label{fig:main}
    \label{fig:Figure-result_ES}
    \end{minipage}
    \hspace{0.01\textwidth}
    \begin{minipage}{0.32\textwidth}
        \centering
        \includegraphics[width=\linewidth]{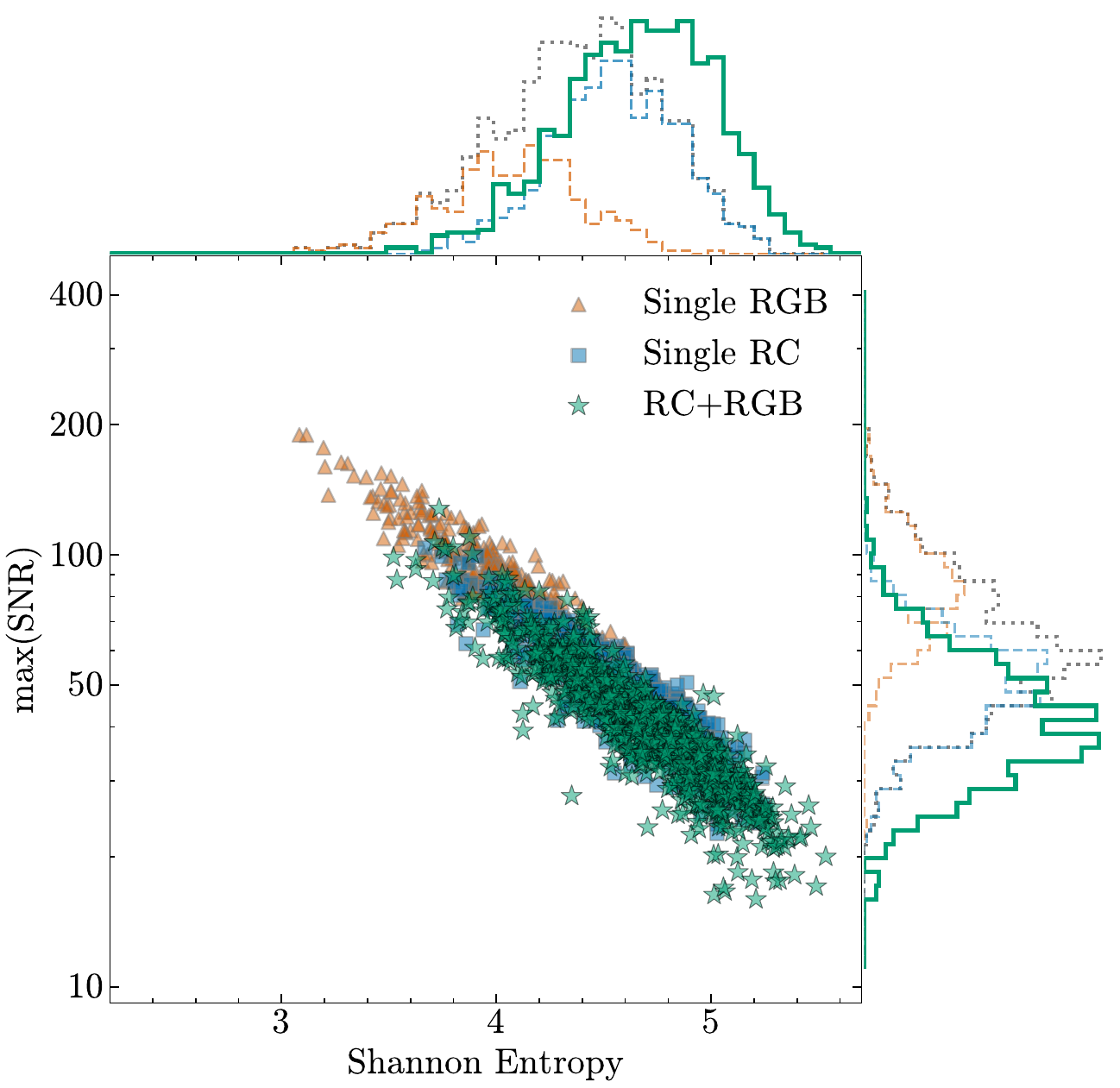}
        \label{fig:RC+RGB}
    \end{minipage}
    \hspace{0.005\textwidth}
    \begin{minipage}{0.32\textwidth}
        \centering
        \includegraphics[width=\linewidth]{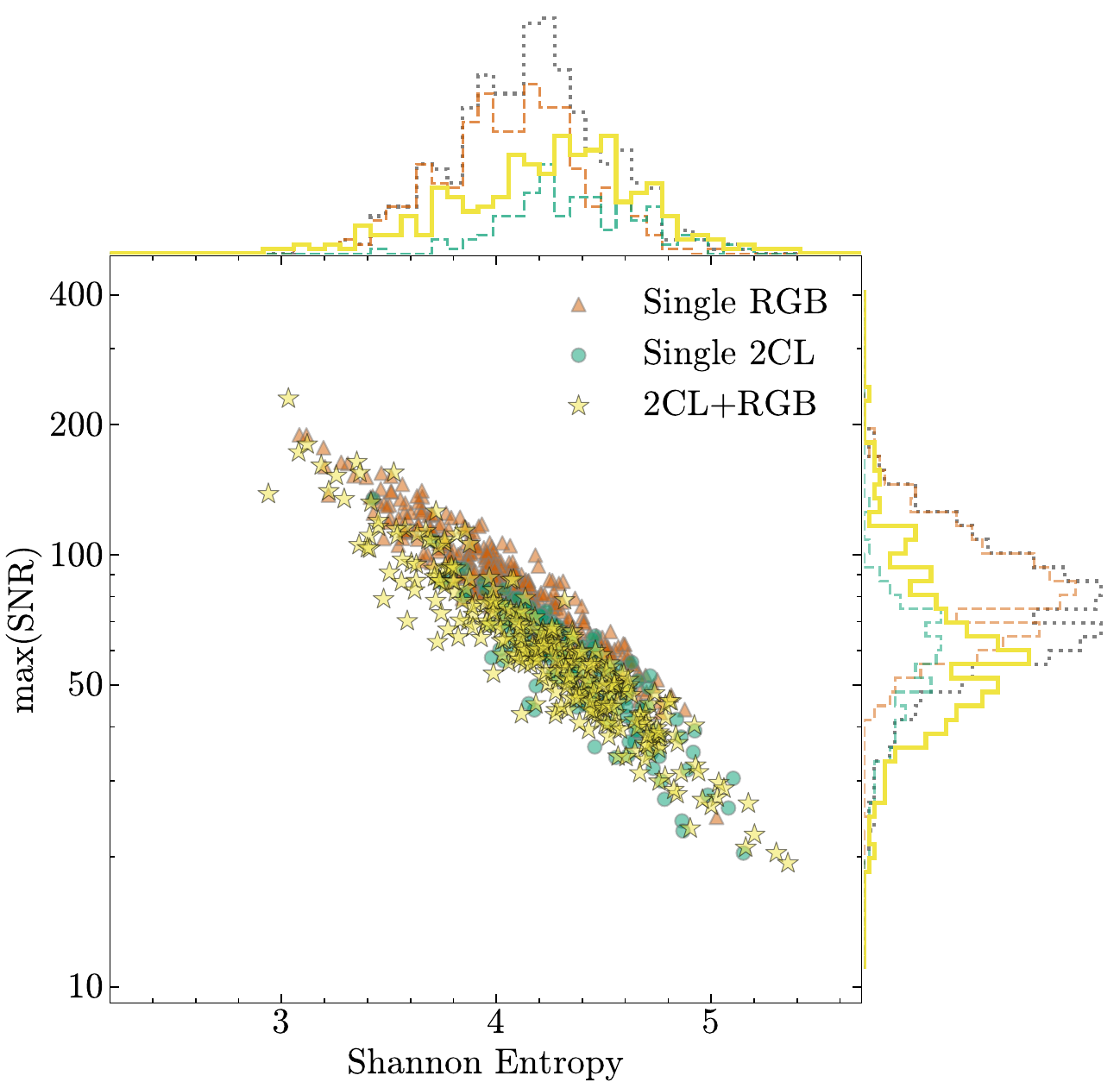}
        \label{fig:2CL+RGB}
    \end{minipage}
\end{figure*}

\begin{figure*}[!htbp]
    \centering
        \includegraphics[width=\linewidth]{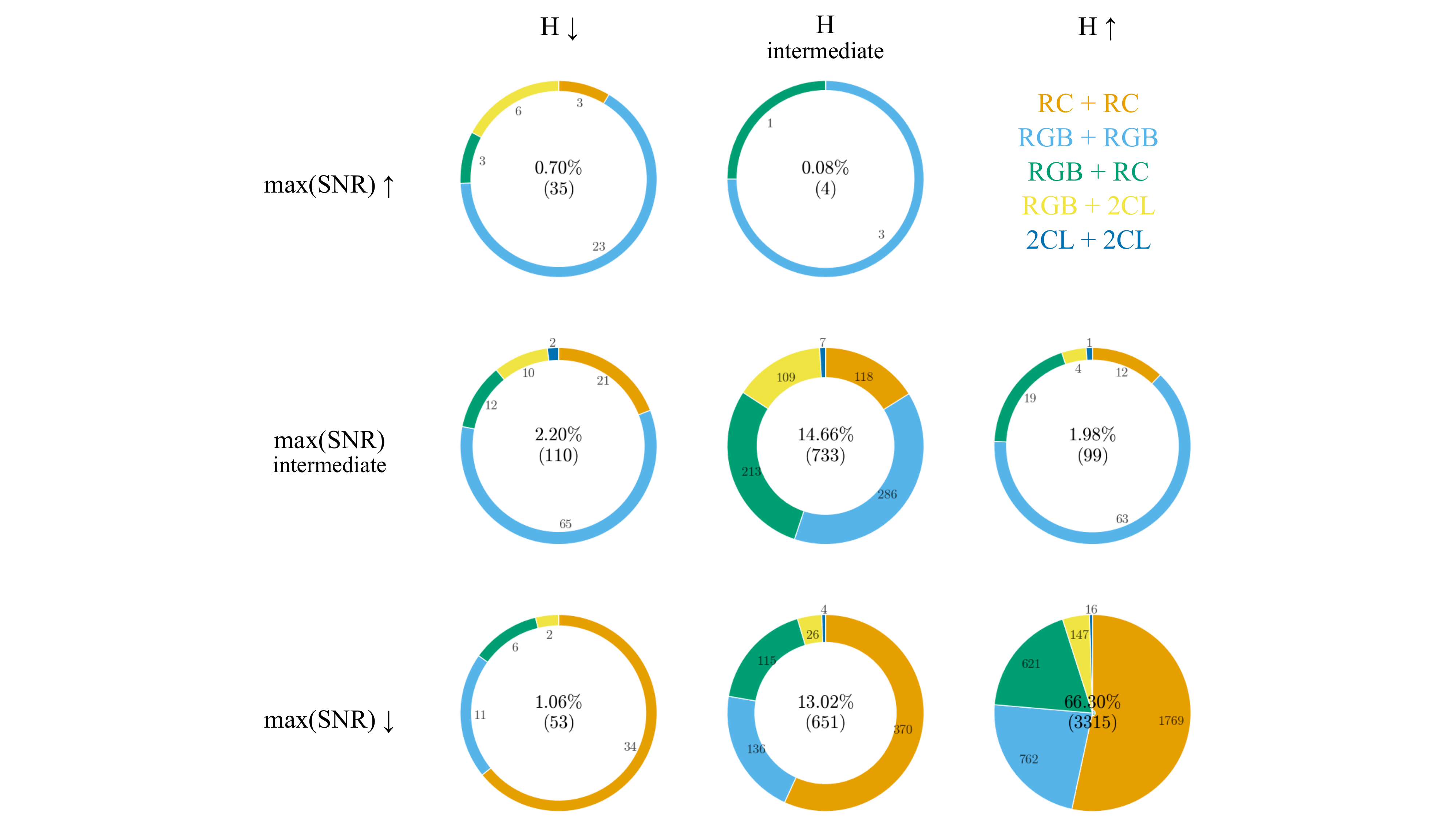}
        \caption{The categorization of the changes in entropy and maximum SNR for AABs compared to their component stars. From left to right, AABs show lower, intermediate, and higher entropy, while from top to bottom, they exhibit higher, intermediate, and lower maximum SNR. The widths of the donut charts indicate the total number of AABs in each category, with larger counts indicated as thicker donuts.}
    \label{fig:Figure-9box}
\end{figure*}

Compared to RGB stars, CHeB stars have larger period spacings of dipole mixed modes and higher number of mixed modes with substantial amplitudes which leads to higher complexity in their PDS and higher entropy. Consequently, AABs composed of two CHeB stars typically result in higher entropy than AABs composed of two RGB stars (see Fig. \ref{fig:Figure-result_ES}). Furthermore, we compare the entropy and maximum SNR of the PDS of AABs with those of their binary components in Fig. \ref{fig:Figure-result_ES}. Most AABs ($\sim$66\%, see Fig. \ref{fig:Figure-9box}) exhibit lower maximum SNR and higher entropy than their components. The reduced oscillation power in artificial binaries is due to the added flux of the two stars reducing their relative flux amplitudes compared to the individual stars (see Fig. \ref{fig:Figure-synthetic}). In Fig. \ref{fig:Figure-9box}, the second most common category is characterized by entropy and maximum SNR values that lie between those of the individual components (comprising about 15\% of AABs). These AABs mostly consist of binary components with significantly different entropy and maximum SNR from each other. In particular, many AABs in this category involve RGB stars, as we discussed in Sect. \ref{subsec:subsec1_Distribution}. On the other hand, only minor differences in entropy and maximum SNR are observed when two CHeB stars form AABs within this category. 

To understand these differences, we calculated the changes in entropy ($\rm{\Delta H =  H_{AAB}-\overline{H}_{stars}}$) and maximum SNR ($\rm{\Delta max(SNR)=max(SNR)_{AAB}-\overline{max(SNR)}_{stars}}$), where $\rm{\overline{H}_{stars}}$ and $\rm{\overline{max(SNR)}_{stars}}$ indicate the mean value. Small differences are observed when one component star dominates the light contribution to the combined signal or both components have oscillation modes located at similar frequencies which lead to oscillation patterns of AABs that resemble the dominant frequency patterns (e.g. the first two columns of Fig. \ref{fig:Figure-RCRCexample}). In contrast, distinctly different oscillation patterns between binary component stars (e.g. last two columns in Fig. \ref{fig:Figure-RCRCexample}) lead to larger differences in both entropy and maximum SNR of the resulting AABs, indicating that the magnitude of these differences depends on the oscillation characteristics of the binary components.

\subsection{Classification of different PDS morphologies of AABs} \label{subsec:subsec3_Classification}

The interplay between brightness, intrinsic oscillation power, complexity and alignment of oscillation patterns determines the PDS of AABs. In particular, the flux ratio mainly constrains whether oscillations from both stars are visible in the PDS. As the flux ratio increases, oscillation signatures from both stars become more clearly detectable. Therefore, we focused on the 394 AABs with flux ratios $\geq$ 0.9 and classified them into the following categories by checking the alignment of their oscillation mode frequencies (see representative examples in Fig \ref{fig:Figure-RCRCexample}, and additional examples in Appendix \ref{sec:appendix}). 

\begin{itemize}
  \item \textbf{Single star-like}, AABs where oscillations from one component dominate due to the stronger intrinsic oscillation power, i.e., the PDS of AABs resembles that of the dominant star with reduced oscillation power. 

  \item \textbf{Aligned}, AABs where the same spherical degrees of oscillation modes of the binary components are located at similar frequencies. 

  \item \textbf{Partially aligned}, AABs where some oscillation modes are located at similar frequencies while others are not. For several cases, a radial mode from one star nearly coincides with a quadrupole mode from the other star, creating a distinct peak in the PDS of artificial binaries. As shown in the third column of Fig. \ref{fig:Figure-RCRCexample}, the quadrupole mode frequencies of the star in the upper panel align with the radial mode frequencies of the star in the middle panel. The mixture of dipole and quadrupole modes complicates the identification of the individual oscillation modes in the AABs. 

  \item \textbf{Misaligned}, AABs where none of the radial or quadrupole modes have similar frequencies between the binary components. Consequently, the PDS of artificial binaries exhibits a highly complex oscillation pattern. Even the radial modes become challenging to be identified. These cases also tend to show more significant variations in entropy and maximum SNR compared to those of the individual binary components.
\end{itemize}

\begin{figure}[!htbp]
    \centering
    \includegraphics[width=0.8\linewidth]{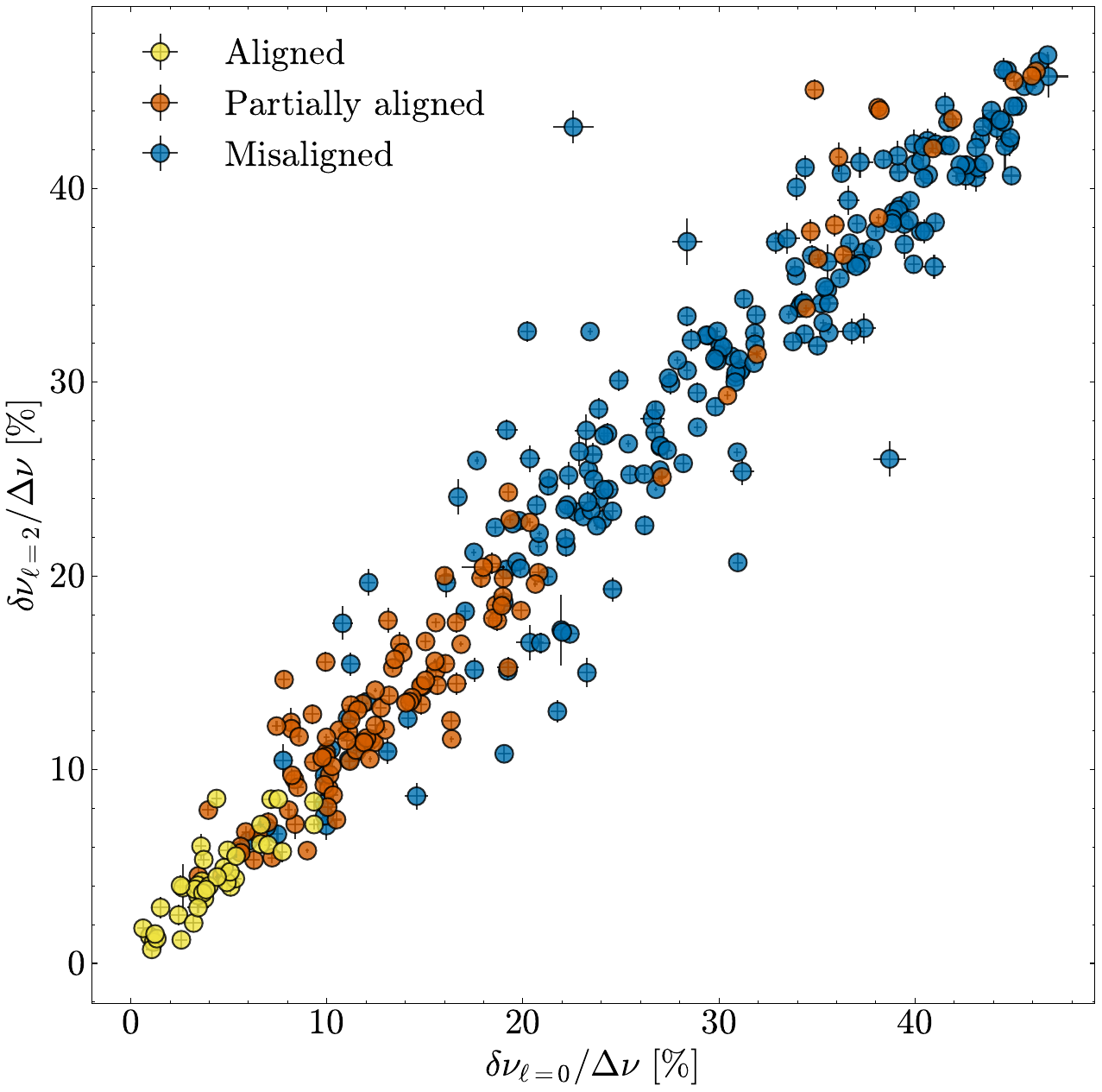}
    \caption{Scaled frequency differences for the radial (\textit{l}=0) versus quadrupole (\textit{l}=2) modes of binary components with flux ratio $\geq$ 0.9. The relative differences ($\delta \nu_{\ell} = |\nu_{\ell}^{(\mathrm{star1})}- \nu_{\ell}^{(\mathrm{star2})}|$) between oscillation modes are normalized by the average $\rm{\Delta\nu}$ of the binary components and expressed as percentages. Data points are color-coded by PDS morphologies of the AABs.}
    \label{fig:Figure-morphologies_frequencies}
\end{figure}

We found that if the frequency differences both in \textit{l}=0 and \textit{l}=2 are less than about 10\% of the mean $\rm{\Delta\nu}$, it typically indicates aligned cases. AABs with partially aligned binary components show frequency differences typically between 10\% and 25\%, and differences greater than 25\% tend to show totally misaligned components (see Fig. \ref{fig:Figure-morphologies_frequencies}). We confirmed these classifications through visual inspection. Importantly, most AABs with flux ratio $\geq$ 0.9 are classified as partially aligned or misaligned (see Table. \ref{tab:classification}). This implies that red-giant asteroseismic binaries whose components have similar oscillation frequencies are likely to show complex features in their PDS when oscillations from both components are clearly detectable. Although the entropy values of AABs do not show clear distinctions between the categories, we found that changes in entropy values are relatively larger in partially aligned and misaligned categories, both with median of $\rm{|\Delta H| \approx 0.4 }$ bits, compared to aligned cases with a median of $\rm{|\Delta H| \approx 0.3}$ bits (see Fig \ref{fig:Figure-appendix_entropy}).

\begin{table}[!htbp]
    \caption{Results of the PDS morphology classification of AABs with flux ratio $\geq$ 0.9.}
    \centering
    \begin{tabular}{c|c}
    \hline\hline
    Type& \#\\
    \hline
    Single star-like & 17\\
    Aligned&  40\\
    Partially aligned & 114\\
    Misaligned & 223\\
    \hline\hline
    \end{tabular}
    \tablefoot{Columns type, and \# list the type of morphologies and the number of AABs. See details in Sect. \ref{subsec:subsec3_Classification}.}
    \label{tab:classification}
\end{table}

\begin{figure*}[!htbp]
    \centering
    \includegraphics[width=\linewidth]{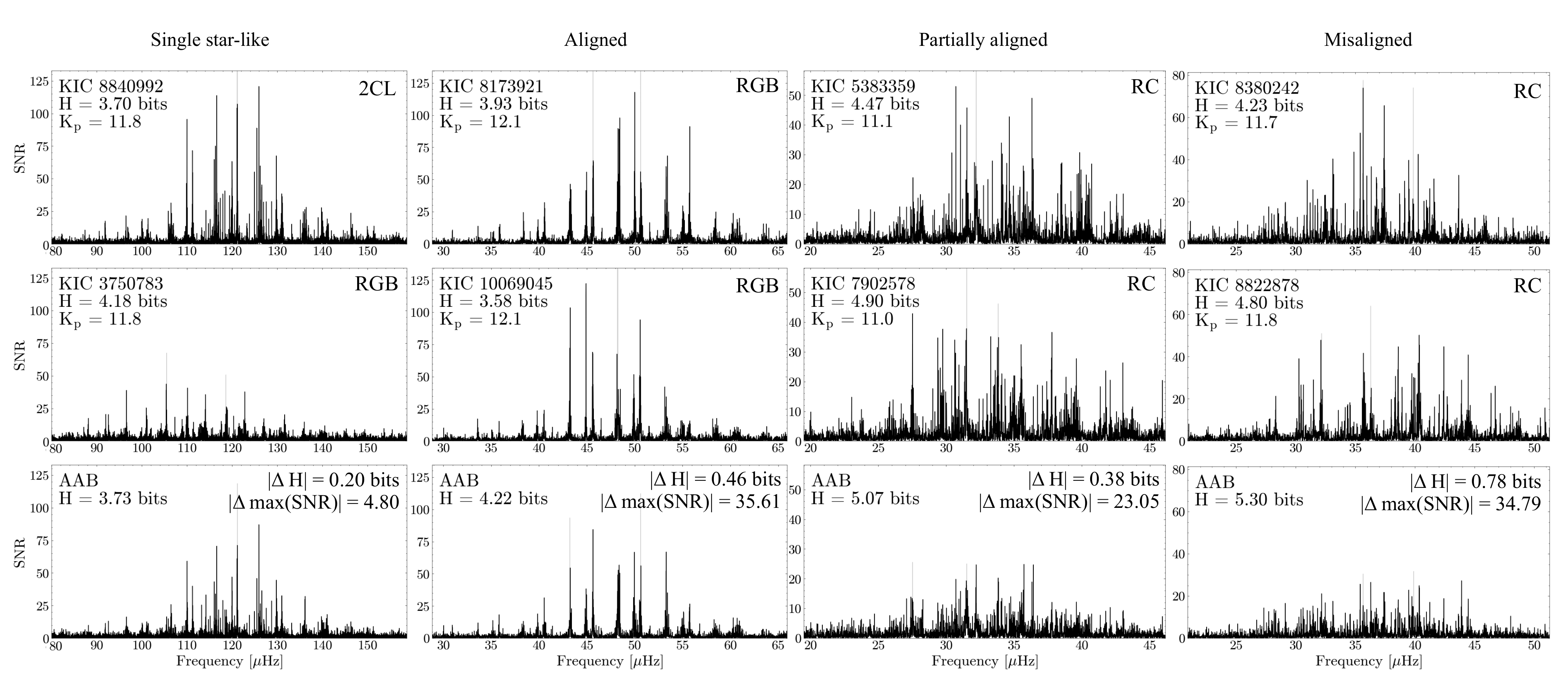}
    \caption{PDS of artificial asteroseismic binaries which represent various morphological types, along with the corresponding PDS of the individual component stars that created them. Gray peaks indicate filtered out peaks in the entropy calculation (see Sect. \ref{subsec:entropy}).}
    \label{fig:Figure-RCRCexample}
\end{figure*}

\section{Discussion and Conclusions} \label{sec:Conclusion}
In this paper, we created 5,000 artificial asteroseismic binaries to explore the characteristics of unresolved red-giant asteroseismic binaries. This extensive set of artificial asteroseismic binary (AAB) systems allowed us to predict observable features and identify specific spectral signatures useful for recognizing seismically unresolved asteroseismic binaries in the observed data. Among 5,000 AABs, about 47\% of the AABs consisted of two red clump stars. Importantly, most of the AABs (about 66\%) exhibited increased Shannon entropy and decreased maximum SNR compared to their individual components. 

The primary goal of our study was to investigate the PDS when oscillations from two red giants are present, particularly when they oscillate at similar frequency ranges. Thus, we focused on AABs created from stars with a flux ratio greater than 0.9 (about 8\% out of 5,000 AABs), where we expected the oscillation signatures from both stars to be detectable. 

We classified the PDS morphologies of AABs (flux ratio $\geq$ 0.9) into four categories: single star-like, aligned, partially aligned, and misaligned. Most of them classified to either the partially aligned or misaligned categories. This suggests that seismically unresolved red-giant asteroseismic binaries oscillating at similar frequency ranges are likely to exhibit complex oscillation patterns in their PDS. 

Interestingly, several observed stars exhibit unusually complex oscillation patterns with low oscillation power. In addition to the scenarios such as prolonged mass-loss events extending from the RGB phase into the CHeB phase (e.g. KIC 9508595; \citealt{2019_Elsworth,2022_Braun}) or extreme coupling between pressure- and gravity-mode cavities (e.g. KIC 11299941; \citealt{2023_Matteuzzi}), we propose a scenario that observed stars with complex PDS may be asteroseismic binaries. 

In conclusion, seismically unresolved asteroseismic binaries offer a explanation to understand the observed stars with complex PDS. Further observational studies are essential to validate our hypothesis and refine the criteria for identifying unresolved red-giant asteroseismic binaries. Comparing our AABs with actual observations will be crucial for confirming these scenarios and will be presented in the future work. 

\begin{acknowledgements}
        We thank the anonymous referee for their constructive feedback that improved the manuscript considerably. We acknowledge funding from the ERC Consolidator Grant DipolarSound (grant agreement \# 101000296). In addition, we acknowledge support from the Klaus Tschira Foundation. This paper includes data collected by the \textit{Kepler} mission. Funding for the \textit{Kepler} mission is provided by the NASA Science Mission Directorate. 
\end{acknowledgements}

\bibliographystyle{aa.bst}
\bibliography{reference}

\begin{appendix} 
\onecolumn
\section{PDS Morphologies of AABs with different evolutionary stage combinations and their entropy analysis} \label{sec:appendix}

\begin{figure*}[h!]
    \centering
    \includegraphics[width=\linewidth]{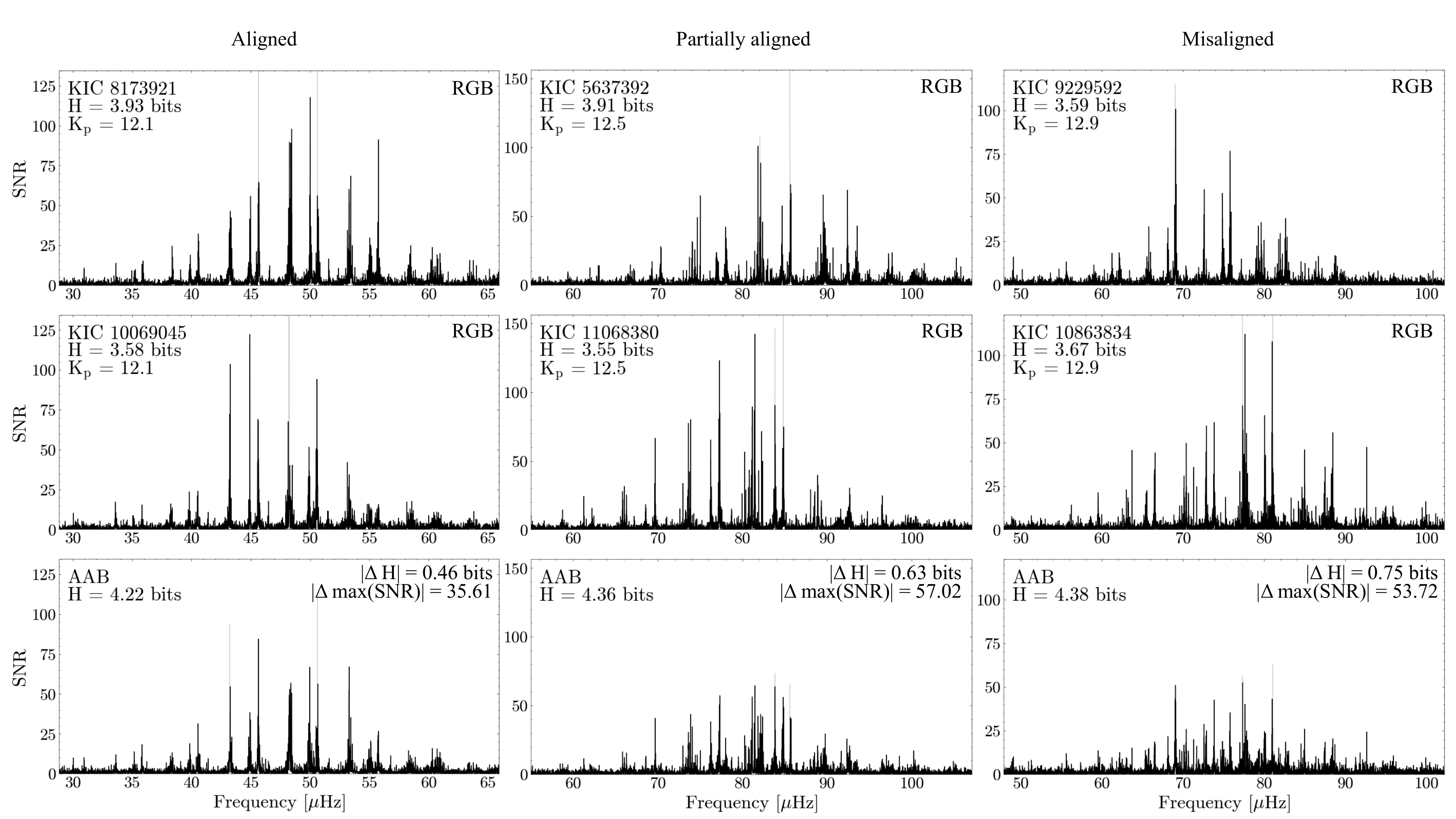}
    \vspace{-0.5cm}
    \caption{Same as Figure \ref{fig:Figure-RCRCexample}, now for the RGB+RGB where oscillation signatures from both stars are clearly detectable.}
    \label{fig:Figure-appendix_RGBRGB}
\end{figure*}

\begin{figure*}[h!]
    \centering
    \includegraphics[width=\linewidth]{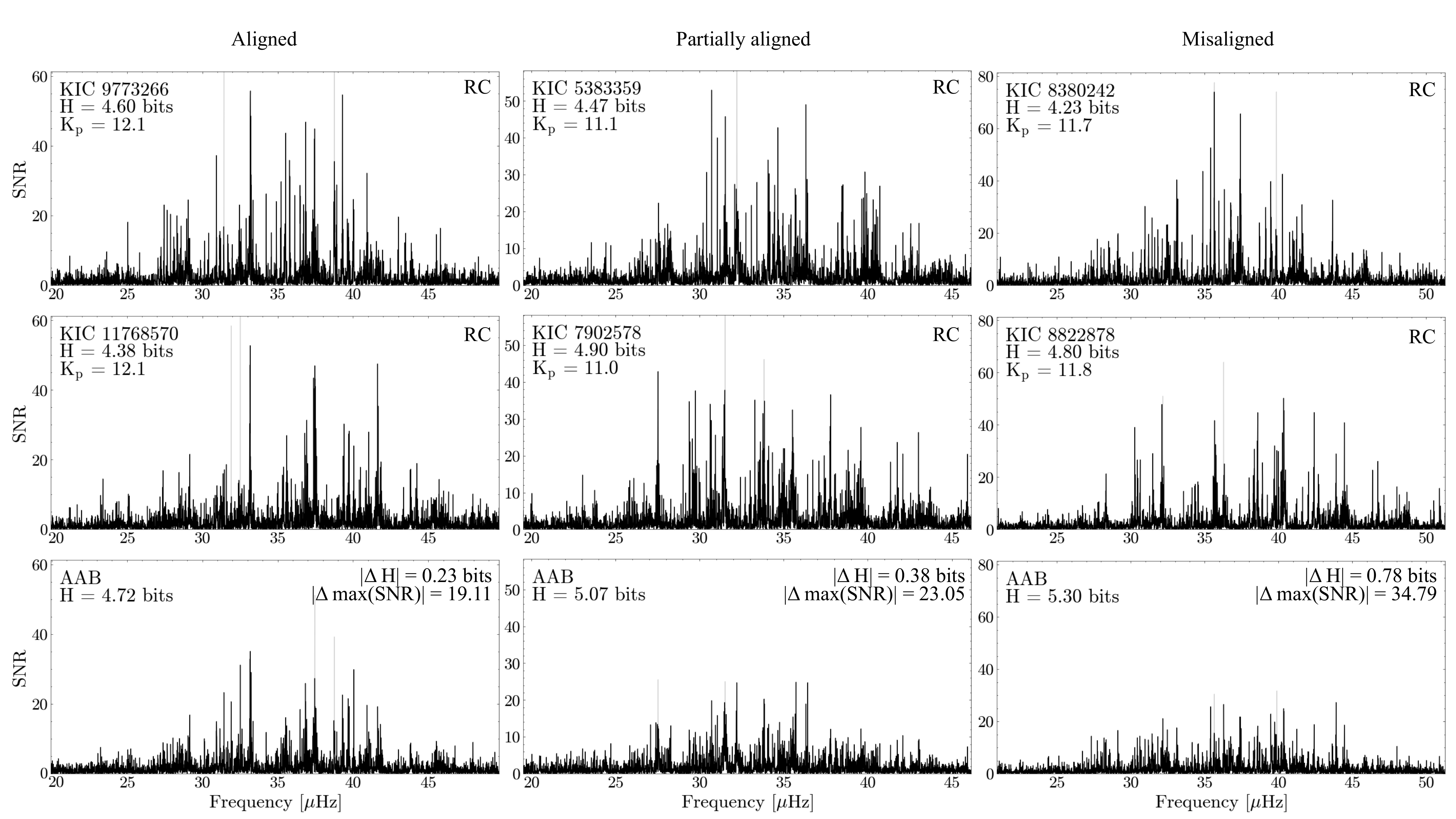}
    \caption{Same as Figure \ref{fig:Figure-appendix_RGBRGB}, now for the RGB+RC}
    \label{fig:Figure-appendix_RCRC}
\end{figure*}

\begin{figure*}[h!]
    \centering
    \includegraphics[width=\linewidth]{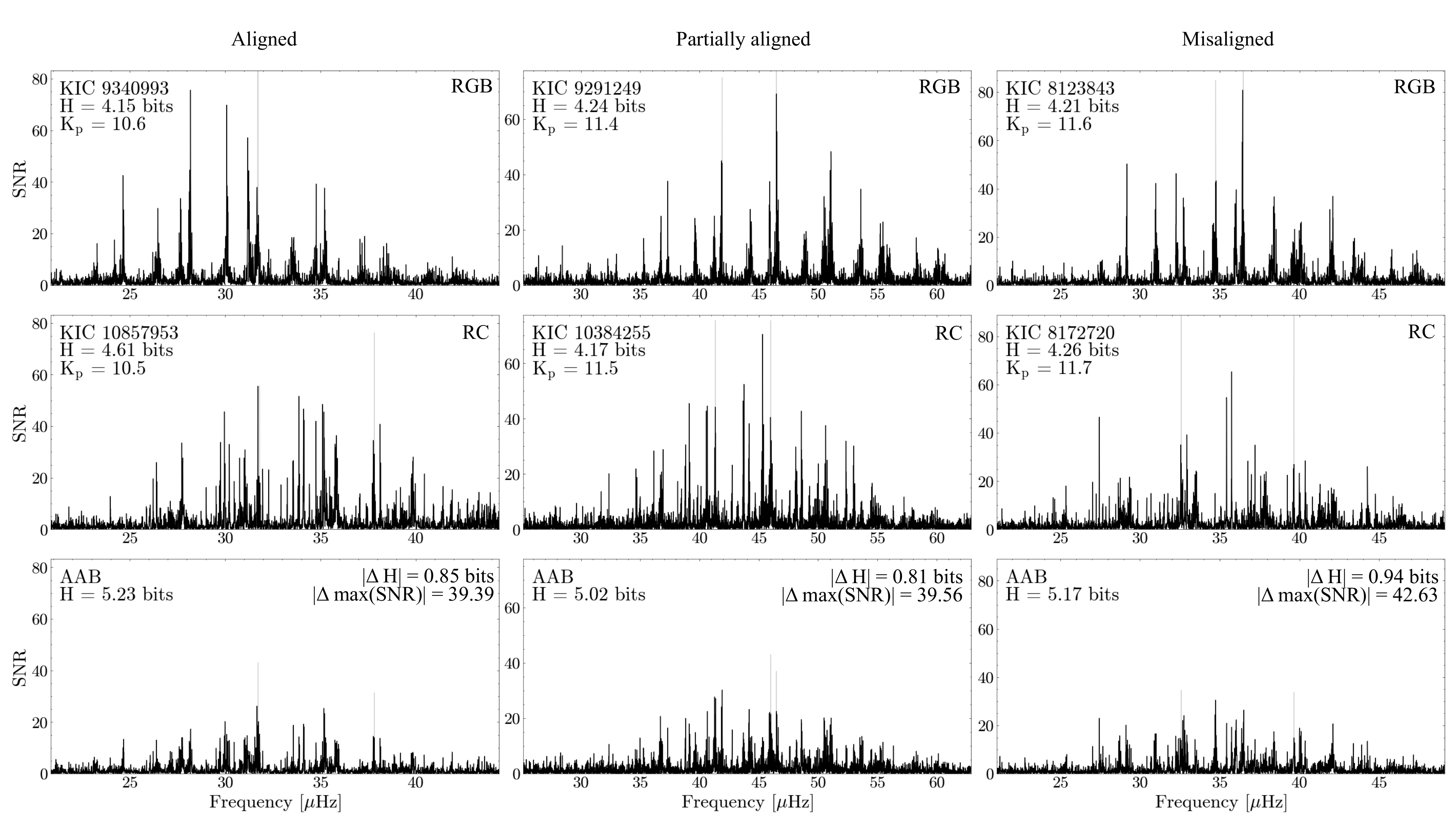}
    \caption{Same as Figure \ref{fig:Figure-appendix_RGBRGB}, now for the RGB+RC}
    \label{fig:Figure-appendix_RGBRC}
\end{figure*}

\begin{figure*}[h!]
    \centering
    \hspace{-0.5cm}
    \includegraphics[width=\linewidth]{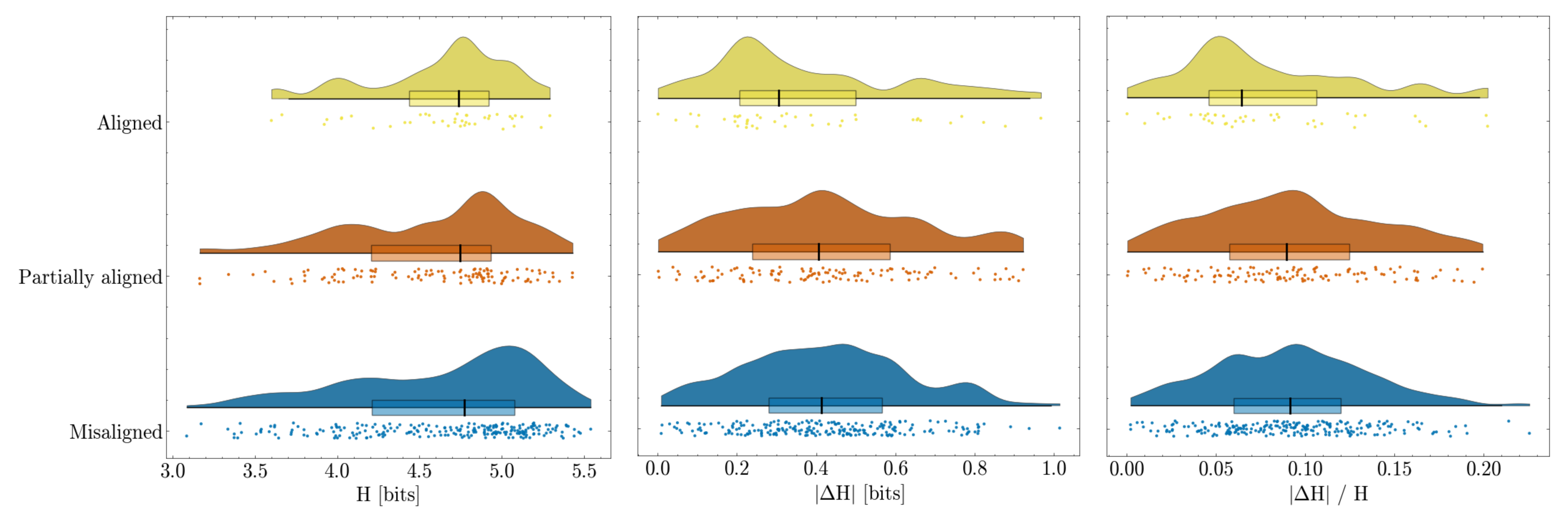}
    \caption{Distributions of (left) the entropy values (H), (middle) the absolute difference in entropy between AABs and the mean entropy of its component stars 
     ($\rm{|\Delta H =  H_{AAB}-\overline{H}_{stars}|}$), and (right) the relative entropy difference ($\rm{|\Delta H|/H}$) for different PDS morphologies of AABs classified in Sect. \ref{subsec:subsec3_Classification}. The figure structure follows that of Fig. \ref{fig:Figure-pairs}.}
    \label{fig:Figure-appendix_entropy}
\end{figure*}

\end{appendix}
        
\end{CJK}
\end{document}